\RequirePackage{lineno}

\documentclass[prd,twocolumn,showpacs,superscriptaddress,amsmath,amssymb,nofootinbib]{revtex4-1}
\usepackage{graphicx}
\usepackage{epsfig}
\usepackage{overpic}
\usepackage{dcolumn}
\usepackage{bm}
\usepackage{color}
\usepackage{lineno}
\usepackage{xspace}
\usepackage{verbatim}
\usepackage{subfigure}
\include{def-com}
\usepackage{hyperref}
\hypersetup{colorlinks,linkcolor={blue},citecolor={blue},urlcolor={red}}
\usepackage{booktabs,tabularx}
\usepackage{array}
\usepackage{titlesec}
\usepackage{multirow}
\newcolumntype{L}[1]{>{\raggedright\let\newline\\\arraybackslash\hspace{0pt}}m{#1}}
\newcolumntype{C}[1]{>{\centering\let\newline\\\arraybackslash\hspace{0pt}}m{#1}}
\newcolumntype{R}[1]{>{\raggedleft\let\newline\\\arraybackslash\hspace{0pt}}m{#1}}
\begin{document}
\title{\boldmath Feasibility study of $CP$ Violation in $\tau^{-}\to K_{S}\pi^{-}\nu_{\tau}$ decays at Super Tau Charm Facility}
\author{
  \begin{small}
    \begin{center}
\boldmath Haoyu Sang$^{1,2}$\, Xiaodong Shi$^{1,2}$\, Xiaorong Zhou$^{1,2}$\, Xianwei Kang$^{3,4}$\, Jianbei Liu$^{1,2}$
\\
\vspace{0.2cm} {\it
$^{1}$ State Key Laboratory of Particle Detection and Electronics, Hefei 230026, People's Republic of China\\
$^{2}$ University of Science and Technology of China, Hefei 230026, People's Republic of China\\
$^{3}$ Key Laboratory of Beam Technology of Ministry of Education, College of Nuclear Science and Technology, Beijing Normal University, Beijing 100875, People's Republic of China\\
$^{4}$ Beijing Radiation Center, Beijing 100875, China
}\end{center}
\vspace{0.4cm}
\end{small}
}
\noaffiliation{}
\begin{abstract}
We report a feasibility study of $CP$ violation of $\tau^{-}\rightarrow K_{S}\pi^{-} \nu_{\tau}$ decay at a Super Tau Charm Facility~(STCF).
With an expected luminosity of 1~ab$^{-1}$ collected by STCF per year at a center-of-mass energy of 4.26~GeV, the statistical sensitivity for the $CP$ violation is determined to be of order
$9.7\times10^{-4}$ by measuring the decay-rate difference between $\tau^{+}\rightarrow K_{S}\pi^{+}\bar{\nu}_{\tau}$ and
$\tau^{-}\rightarrow K_{S}\pi^{-} \nu_{\tau}$.  The analysis is performed using a reliable fast simulation software package,
which can describe the detector responses properly and vary the responses flexibly for further optimization. 
Moreover, the energy-dependent efficiencies for reconstructing $\tau^{-}\rightarrow K_{S}\pi^{-} \nu_{\tau}$ are presented 
and the expected $CP$ sensitivity is proportional to $1/\sqrt{\mathcal{L}}$ in the energy region from 4.0 to 5.0~GeV. 
The sensitivity of $CP$ violation is of order $3.1\times10^{-4}$ with 10~ab$^{-1}$ integrated luminosity, which is equivalent 
to ten years data taking in this energy region at STCF.
\end{abstract}
\maketitle
\section{\boldmath INTRODUCTION}

The Cabibbo-Kobayashi-Maskawa~(CKM) matrix can accommodate $CP$ violation~(CPV)
in a natural way with a complex phase in the quark mixing sector~\cite{CKM1,CKM2}.
Experimentally, the CPV has been observed in the meson sector, firstly in $K$ meson~\cite{kaoncp1,kaoncp2,kaoncp3},
subsequently in $B$ meson~\cite{Bcp1,Bcp2,Bcp3,Bcp4,Bcp5,Bcp6} and most recently in charm meson decays~\cite{charmcp},
and all results to date are consistent with the predictions of the CKM mechanism in the Standard Model~(SM).

However, the origin of CPV has remained as an unsolved problem as it is not clear
if CKM mechanism is the unique source. The matter-antimatter asymmetry of the universe~\cite{matterasy1,matterasy2}
indicates that there must be non-SM CPV sources, and such additional sources are indeed incorporated in
many extensions of the SM~\cite{extendSM1,extendSM2}. Currently, CPV has not been observed in the
lepton sector, and any significant observation would be a clear indication of physics beyond the SM. In an intriguing scenario~\cite{Yanagida}, the baryogenesis is argued to be mainly driven by the leptogenesis, and then CPV is required in leptodynamics.
Thus exploration of CPV in lepton sector does provide a different and complementary landscape, at least.
Among the lepton sector, $\tau$ decay is a good place to seek for CPV either within or beyond the SM~\cite{taucp1,taucp2,taucp3,BellePhysicsBook,Pich},
since it has abundant hadronic decay channels with sizable branching ratios, and half part of the matrix element is purely electroweak and
the involved hadrons are generated by quark-antiquark pairs, such that they are factorized, exempting from complicacy of pure hadronic decay.

In this paper, we will discuss the specific channel $\tau^{-}\to K_{S}\pi^{-}\nu_{\tau}$ and concentrate on 
the CPV and related New Physics~(NP) searches. This channel is also of great importance
for the measurements of form factors and extraction of $|V_{us}|$ element in CKM matrix, and has rich intermediate
resonances which provides an important play-ground for non-perturbative QCD study. 
Experimentally, a rich source of $\tau$ leptons comes from electron-positron collider
and the physics programs related with $\tau$ decays are usually performed at B-factory or Z-factory. There
are a lot prospects concern the potential of $\tau$ decays at future factories of such~\cite{belleii}.
However, the $\tau$ decays are rarely investigated in tau-charm energy region due to the low statistics available  and difficulty in reconstruction. 
A feasibility study of $\tau$ decays in tau-charm region will be of great importance to complete the global picture of whole $\tau$ physics.
The proposed STCF~\cite{stcf} in  China is designed to have a peaking luminosity of $>0.5\times10^{35}$~cm$^{-2}$s$^{-1}$ at $\sqrt{s}=4$~GeV.
It is a symmetry electron-positron beam collider designed to provide  $e^{+}e^{-}$ interactions at a
c.m.~energy $\sqrt{s}$ from 2.0 to 7.0 GeV. 
The $\tau$ leptons are selected via process $e^{+}e^{-}\to\tau^{+}\tau^{-}$ where the cross section of  $e^{+}e^{-}\to\tau^{+}\tau^{-}$
is peaking around $\sqrt{s}=$4.26~GeV, to be 3.5~nb~\cite{tsai}.
The number of $\tau$ pairs produced at STCF is estimated to be $3.5\times10^{9}$ with an expected integrated luminosity of 1~ab$^{-1}$ per year.

This paper is organized as follows: In Sec.~2, we elaborate the physical significance for investigating the channel $\tau^{-}\to K_{S}\pi^{-}\nu_{\tau}$.
In Sec.~3, the detector concept for STCF is introduced as well as the Monte Carlo~(MC) samples used for this study. 
The event selection for the signal process and optimization of detector optimization are presented in Sec.~4 and Sec.~5.
Sec.~6 is the discussion of the results and Sec.~7 is the conclusion.

\section{\boldmath PHYSICS PROGRAM OF $\tau^{-}\to K_{S}\pi^{-}\nu_{\tau}$}

Within the SM, there is no direct CPV in hadronic $\tau$ decays at the tree level in weak
interaction, however, the well-measured CPV  in $K_{L}\to\pi^{\mp}l^{\pm}\nu$~$(l=e,\mu)$ produces a difference in 
$\Gamma(\tau^{+}\to K_{L}\pi^{+}\bar{\nu}_{\tau})$
vs. $\Gamma(\tau^{-}\to K_{L}\pi^{-}{\nu_{\tau}})$ due to the $K^0-\bar K^0$ oscillation.
The same asymmetry also appears in $\Gamma(\tau^{+}\to K_{S}\pi^{+}\bar{\nu}_{\tau})$ vs. $\Gamma(\tau^{-}\to K_{S}\pi^{-}{\nu_{\tau}})$, and is calculated to be~\cite{taucp5,taucp6}:
\begin{equation}\label{ACPTh}
\begin{split}
&A_{CP}(\tau^{-}\to K_{S}\pi^{-}\nu_{\tau})\\&\qquad=\frac{\Gamma(\tau^{+}\to K_{S}\pi^{+}\bar\nu_{\tau})-\Gamma(\tau^{-}\to K_{S}\pi^{-}\nu_{\tau})}{\Gamma(\tau^{+}\to K_{S}\pi^{+}\bar\nu_{\tau})+\Gamma(\tau^{-}\to K_{S}\pi^{-}\nu_{\tau})} \\
 & \qquad =(0.33\pm0.01)\%.
\end{split}
\end{equation}

Experimentally, BaBar experiment has found evidence for CPV in $\tau$ decays~\cite{CPbabar}:
\begin{equation}
\label{expcp}
A_{CP}(\tau^{-}\to K_{S}\pi^{-}\nu_{\tau}[\geq 0\pi^{0}])=(-0.36\pm0.23\pm0.11)\%,
\end{equation}
and there is a $2.8$ standard deviation difference from the theoretical prediction in Eq.~\eqref{ACPTh}. 
It should be noted that, the $K^0-\bar K^0$ oscillation leads
to the same value of CPV for inclusion of any $\pi^0$s in $\tau$ decays from SM prediction.

Motivated by the above CPV disagreement, various NP scenarios are proposed, {\it e.g.}, introducing a non-standard tensor interaction \cite{tensor1, tensor2}. 
In Ref.~\cite{tensor2}, a very large angular-weighted CPV is predicted additionally with such tensor interaction. 
However, those conclusions are questioned in Ref.~\cite{Martin}, where the authors point out the tensor contributions used in Refs.~\cite{tensor1, tensor2} are overestimated due to lacking of the well-known Watson theorem of final-state interaction, resulting in very large values of the imaginary parts
of the Wilson coefficients to explain the $CP$ anomaly; those large values of Wilson coefficients contradict with the bound from
neutron electric dipole moment (EDM) and $D-\bar D$ mixing (unless NP occurs below the electroweak breaking scale). The standpoint of Ref.~\cite{Martin} is
adopted later in Refs.~\cite{RoigPRD, XinQiangLi}, and especially, Ref.~\cite{XinQiangLi} finds that once imposing the combined constraints from
the branching ratio and decay
spectrum, it is hard to explain the experimental values of CPV. 
But again, Ref.~\cite{Martin} is questioned by the authors of Ref.~\cite{Roy}---they point out that the tensor operator used in Ref.~\cite{Martin} corresponds to only a specific way of imposing gauge invariance, and instead, introducing new type of tensor operator could account for the $CP$ anomaly while
evading bounds from the neutron EDM and keeping
the extraction of $|V_{us}|$ from exclusive $\tau$ decay unaffected. In brief,  BaBar's CPV observables leads to a series of theoretical discussions,
and those theoretical discussions are in the state of confusion. Therefore, a new measurement on the experimental side is essential to approach any
conclusive statement on the NP signal. We stress the importance of precision here. 

Moreover, as discussed in Ref.~\cite{taucp2}, probing $A_{CP}(\tau^{-}\to [K\pi]^{-}\nu_{\tau})$, $A_{CP}(\tau^{-}\to [K 2\pi]^{-}\nu_{\tau})$ and  $A_{CP}(\tau^{-}\to [K3\pi]^{-}\nu_{\tau})$ separately can help to establish the existence of new dynamics since the global CPV as expressed in Eq.~(\ref{expcp}) are often much reduced,
and one needs to understand the basis of the observed data on $\tau^{-}\to K_{S}\pi^{-}\nu_{\tau}[\geq 0\pi^{0}]$ vs. $\tau^{-}\to [K\pi]^{-}\nu_{\tau}$, $\tau^{-}\to [K 2\pi]^{-}\nu_{\tau}$ and  $\tau^{-}\to [K3\pi]^{-}\nu_{\tau}$. In SM, the same CPV holds for the decay including any number of $\pi^0$ meson, due to the sole source by
$K^0-\bar K^0$ oscillation. 
But the situation will change in the presence of NP, {\it e.g.} the two-Higgs Doublet Model has been used to show its influence on 
the different observable sets for the channels $\tau^{-}\to K \pi^{-} \pi^{0} \nu_\tau$ and
$\tau^{-}\to K \pi^{-} \nu_\tau$~\cite{2HDM,Kimura}. Certainly, more complicated impact of resonances appear in the multibody channels, and more statistic data is needed
in this case. We also note that in the four- and five-body decays, one can also access to the CPV by T-odd observables \cite{KangTodd,Kang1,Kang2}.

Apart from the above BaBar measurement, CLEO~\cite{CPcleo} and Belle~\cite{CPbelle} experiments have also focused on the CPV that could arise from a charged scalar boson exchange~\cite{theocp}. This type of CPV can be detected by measuring the $\tau^{\pm}$ decay angular distributions. The CPV is found to be compatible with zero with a precision of $\mathcal{O}(10^{-3})$~\cite{CPbelle}.
However, current experimental sensitivity cannot make a conclusion on the CPV from $\tau$ decay due to large uncertainty. 
Therefore, a higher-precision result is highly required for hunting signals of NP.
At BelleII, the CPV is studied in the angular distribution of $\tau^{-}\to K_{S}\pi^{-}\nu_{\tau}$ decays and 
the precision is expected to be of $\sqrt{70}$ times improved with a 50~ab$^{-1}$ data sample~\cite{CPbelleii}.


In order to capture the faint NP signal, or instead, to constrain the NP model, the accurate knowledge of form factor is also an essential input~\cite{tensor2,RoigPRD}. 
The process $\tau^{-}\to K_s\pi^{-}\nu_{\tau}$ decay in the SM is described by the vector and scalar form factors. 
The vector form factor receives mainly the contribution of $K^{*}(892)$, while $K^{*}(1410)$ is needed in the higher-energy region tail.
For scalar form factor, there is no clear dominance of single resonance. 
Currently the most sophisticated theory description of the form factors are obtained
by the model-independent dispersive representation imposing constraints from chiral symmetry and their asymptotic QCD behavior~\cite{Bernard, Roig2014,Oller}. 
However, other descriptions are available~\cite{Kimura,Belle2007}. In Ref.~\cite{Kimura}, the form factors are
calculated up to one-loop level using the chiral Lagrangian. The resulting shape differs from the dispersive one in Ref.~\cite{Bernard}. 
Such difference may be attributed to the perturbative vs. non-perturbative treatment. In Ref.~\cite{Belle2007}, experimentalist aims to achieve a satisfactory description of data in a phenomenological way, and then a superposition of Breit-Wigner functions is used.
 As already commented in Ref.~\cite{Martin}, the resulting phase does not vanish at threshold and also violates the Watson theorem before the inelasticity sets in. In short, the shapes of $P$-wave form factor agree while the $S$-wave shapes can differ dramatically in different model calculations. However, from a pragmatic point of view, all these variants of form factor could describe the data for the total mass
spectrum almost equally well. A partial-wave analysis
will certainly help to pin down this issue, which again needs more high-quality data. This point has been noted by
Belle collaboration \cite{Belle2007}: the future more data on the invariant mass spectrum of $K_s\pi$ combined with the angular analysis will elucidate the nature of scalar form factor and check various theoretical approaches.
At STCF, high statistics of $\tau^{-}\to K_{S}\pi^{-}\nu_{\tau}$ can put strong constraints  for
the vector and scale form factors. Meanwhile, the precision in the determination of the form factor parameters can be improved.

On the other hand, the $|V_{us}|$ can be extracted from the measurement of the $\tau^{-}\to K_{S}\pi^{-}\nu_{\tau}$ decay. Indeed, the
decay rate  of  $\tau^{-}\to K_{S}\pi^{-}\nu_{\tau}$ can be expressed as~\cite{Vusforkspi}:
\begin{equation}
\label{Vus}
\Gamma = G^{2}_{F}NC^{2}_{K,\tau}S_{EW}(|V_{us}|f^{K^{0}\pi^{-}}_{+}(0))^{2}I_{K}(1+\delta_{EM}+\delta_{SU(2)})^{2},
\end{equation}
where $N$ is normalization coefficient($N=m^{3}_{\tau}/(48\pi^{3})$), $G_{F}$ is the Fermi constant, $C_{K,\tau}$ is Clebsch-Gordan coefficient, $I_{K}$ is the phase space integrals, $S_{EW}$ is the electroweak short-distance, $\delta_{EM}$ is the electromagnetic long-distance, $\delta_{SU(2)}$ is the isospin breaking corrections and $f^{K^{0}\pi^{-}}_{+}(0)$ is the form factor at zero momentum transfer.
According to the above equation, the measurement of $|V_{us}|$ requires: i) accurate measurement of $\Gamma$, ii) precise calculation of $I_{K}$, iii) a good knowledge of the radiative corrections, which are $S_{EW}$, $\delta_{EM}$ and $\delta_{SU(2)}$, iv) a determination of the value of $f^{K^{0}\pi^{-}}_{+}(0)$. In general, with higher precision for $\tau^{-}\to K_{S}\pi^{-}\nu_{\tau}$ branch fraction measurement, the error on $|V_{us}|$ will be reduced~\cite{Vusforkspi2}. 

In short, the physical importance of measuring  $\tau^{-}\to  K_{S}\pi^{-}\nu_{\tau}$ is discussed from three aspects:
i) probing the CPV of this process and testing several NP scenarios 
ii) the resonance parameters (mass and width) can be also determined, which itself is an important topic of the hadron spectrum, 
with assisting us to understand more on the strong impact of intermediate resonances and their interference. We also notice that the large strong phase will be rendered in the resonance region, which together with the weak phase is a requisite for generating large CPV.
iii) the $|V_{us}|$ element can also be determined, which is worth to do, at least as another cross check for the extraction from exclusive decay modes, and for a review see Ref.~\cite{BellePhysicsBook}.
As has been stressed, to make a conclusive remark the precision in the measurement is an essential ingredient. We indeed need to enter the era of high precision.
A detailed analysis of points ii) and iii) will be presented in a future study. In this paper, we
will investigate the reconstruction strategy for $\tau$ decays at STCF,
and will focus on the sensitivity estimation of CPV by measuring the decay rates difference 
between $\tau^{+}\rightarrow K_{S}\pi^{+}\bar{\nu}_{\tau}$ and $\tau^{-}\rightarrow K_{S}\pi^{-} \nu_{\tau}$.
Moreover, to make the best of the large statistic, it is necessary to have a compatibly sophisticated detector at STCF 
to precisely detect and measure particles, where the process $\tau^{-}\to K_{S}\pi^{-}\nu_{\tau}$
 can serve as a benchmark physics process to 
offer constrains to the detector design.

\section{\boldmath DETECTOR  AND MC SIMULATION}
The STCF detector is a general purpose detector designed for $e^{+}e^{-}$ collider which includes a tracking systems composed of the inner and outer trackers,
 a particle identification~(PID) system with $3\sigma$ charged $K/\pi$ separation up to 2~GeV/c, an electromagnetic calorimeter~(EMC) with an excellent
energy resolution and a good time resolution,
a super-conducting solenoid and a muon detector~(MUD) that provides good charged $\pi/\mu$ separation.
The detailed conceptual design for each sub-detector can be found in Ref.~\cite{fastsimu}.
Currently, the STCF detector and the corresponding offline software system are under research and development~(R\&D), 
and it is necessary to have a reliable simulation tool which can access the physics reaches.
A fast simulation tool for STCF has been developed~\cite{fastsimu}, which takes the most common event generator as input to perform a fast and realistic simulation. The simulation includes resolution and
efficiency responses for tracking of final state particles, PID system and kinematic fit related variables.
Besides, the fast simulation also provides flexibly interface for adjusting performance of each sub-system which can be used to optimize the
detector design according to physical requirements.

The process $\tau^{-}\to K_{s}\pi^{-}\nu_{\tau}$ is studied under 1~ab$^{-1}$ integrated luminosity at $\sqrt{s}=4.26$~GeV,
where the cross section of $e^{+}e^{-}\to \tau^{+}\tau^{-}$ is the largest. Unless specified, the charge conjugate decays are always
implied throughout the analysis.
The MC events for $e^{+}e^{-}\to l^{+}l^{-}~(l=e,\mu)$ and $e^{+}e^{-}\to\gamma\gamma$ are generated with {\sc Babayaga}~\cite{babayaga},
and for hadronic production processes~$(e^{+}e^{-}\to q\bar{q})~(q=u,d,s,c)$ are generated with {\sc LundArlw}~\cite{lund}.
The $e^{+}e^{-}\to \tau^{+}\tau^{-}$ process is generated with {\sc KKMC}~\cite{KKMC}, which implements the {\sc Tauola} to 
describe $\tau$ decays inclusively. 
All the MC samples described above are generated according to the expected amount of 1~ab$^{-1}$ integrated luminosity and
no CPV is assigned, and passage of the particles through the detector 
is simulated by the fast simulation software~\cite{fastsimu}.

The signal process $\tau^{-}\to K_{S}\pi^{-}\nu_{\tau}$
are generated with vector and scalar configurations of $K_{S}\pi^{-}$ and the parameterized spectrum 
is described by:
\begin{eqnarray}
\label{equation3}
\frac{d\Gamma}{d\sqrt{s}} \propto && \frac{1}{s}\left(1-\frac{s}{m^{2}_{\tau}}\right)^{2}\left(1+\frac{2s}{m^{2}_{\tau}}\right)P(s)\\ \nonumber
				     &&\times \left\{P^{2}(s)\vert F_{V}\vert^{2}+\frac{3(m^{2}_{K_{S}}-m^{2}_{\pi})^{2}\vert F_{S}\vert^{2}}{4s(1+\frac{2s}{m^{2}_{\tau}})}\right\},
\end{eqnarray}
where $s$ is the squared invariant mass of $K_{S}\pi^{-}$, $m_{\tau}, m_{K_{S}}$ and $m_{\pi}$ are the masses of $\tau, K_{S}$ and charged $\pi^{-}$
from PDG~\cite{pdg}. $P(s)$ is momentum of $K_{S}$  in the ($K_{S}\pi^{-}$) c.m.~frame, given by:
\begin{equation}
P(s) = \frac{\sqrt{(s-(m_{K_{S}}+m_{\pi})^{2})(s-(m_{K_{S}}-m_{\pi})^{2})}}{2\sqrt{s}}.
\end{equation}

The $F_{S}$ and $F_{V}$ are the scalar and vector form factors to parameterize the amplitudes of
$K^{*}_{0}(800)$, $K^{*}(892)$ and $K^{*}(1410)$:
\begin{equation}
F_{S} = a_{K^{*}_{0}(800)}\cdot BW_{K^{*}_{0}(800)},
\end{equation}
\begin{equation}
F_{V} = \frac{BW_{K^{*}(892)}+a_{K^{*}(1410)}\cdot BW_{K^{*}(1410)}}{1+a_{K^{*}(1410)}},
\end{equation}
where BW denotes Breit-Wigner function, and $a_{K^{*}_{0}(800)}$ and $a_{K^{*}(1410)}$ are complex coefficients for the fractions of the $K^{*}_{0}(800)$
and $K^{*}(1410)$ resonances as presented in Ref.~\cite{belleamp}.

\section{\boldmath EVENT SELECTION AND ANALYSIS}

\begin{figure}[htbp]
\begin{center}
\begin{overpic}[width=5.5cm,angle=0]{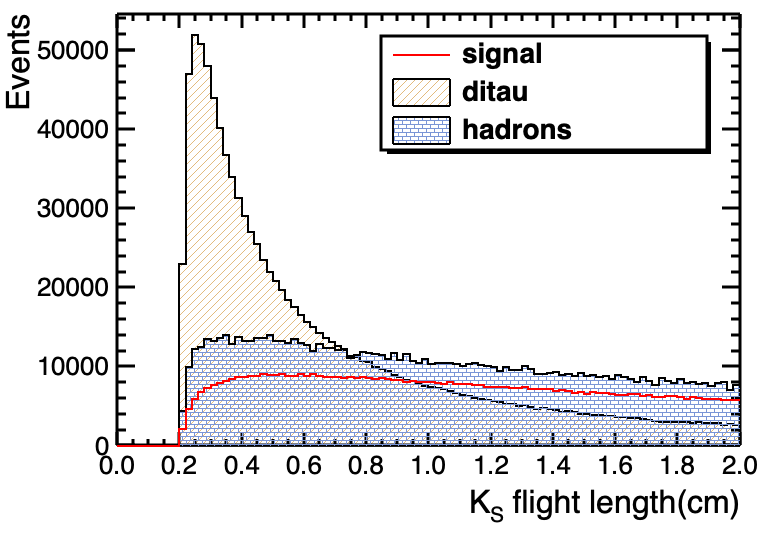}
\end{overpic}
\end{center}
\caption{The flight length of $K_{S}$ candidates for signal and background, where the solid line in red is signal process,
squared shadowed area in blue is  hadronic final states $e^{+}e^{-}\to q\bar{q}$ (hadrons), and
slant shadowed area in yellow is  other $\tau$ decay channels in $e^{+}e^{-}\to\tau^{+}\tau^{-}$ excluding signal process~(ditau), respectively.
MC samples are normalized to 1~ab$^{-1}$ integrated luminosity and all other selection criteria have been applied.}
\label{ksflight}
\end{figure}

The signal events is selected with $\tau^{+}$ decay to leptons, $\tau^{+}\to l^{+}\nu_{l}\bar{\nu}_{\tau}~(l=e,\mu)$,
denoted as tag side, and $\tau^{-}\to K_{S}\pi^{-}\nu_{\tau}$ with $K_{S}\to\pi^{+}\pi^{-}$, denoted as signal side.
At least four charged tracks are required in one event after passing fast simulation, with only the requirement of acceptance applied,
{\it e.g.}$ |\cos\theta|<0.93$, where $\theta$ is the polar angle with respect to the beam direction.
The $K_{S}$ candidates are selected from pairs of oppositely charged tracks, which satisfy a vertex-constrained fit
to a common point. The two charged tracks with minimum $\chi^{2}$ of vertex fit
are assumed to be pions produced from $K_{S}$. The $K_{S}$ is required to have an invariant mass in range $0.485<M_{\pi^{+}\pi^{-}}<0.512$~GeV/$c^{2}$.
Furthermore, considering the finite decay length of $K_{S}$~\cite{pdg},
the significance of flight length of $K_{S}$ candidates is required to be larger than 2 and
the flight length of  $K_{S}$ should be larger than 0.5~cm as shown in Fig.~\ref{ksflight}. 
The background processes with the same final states as signal process 
but contains no $K_{S}$, such as $\tau^{-}\to\pi^{+}\pi^{-}\pi^{-}\nu_{\tau}$, 
can be significantly suppressed by above selections.

\begin{figure}[htbp]
\begin{center}
\begin{overpic}[width=5.5cm, height=5.5cm, angle=0]{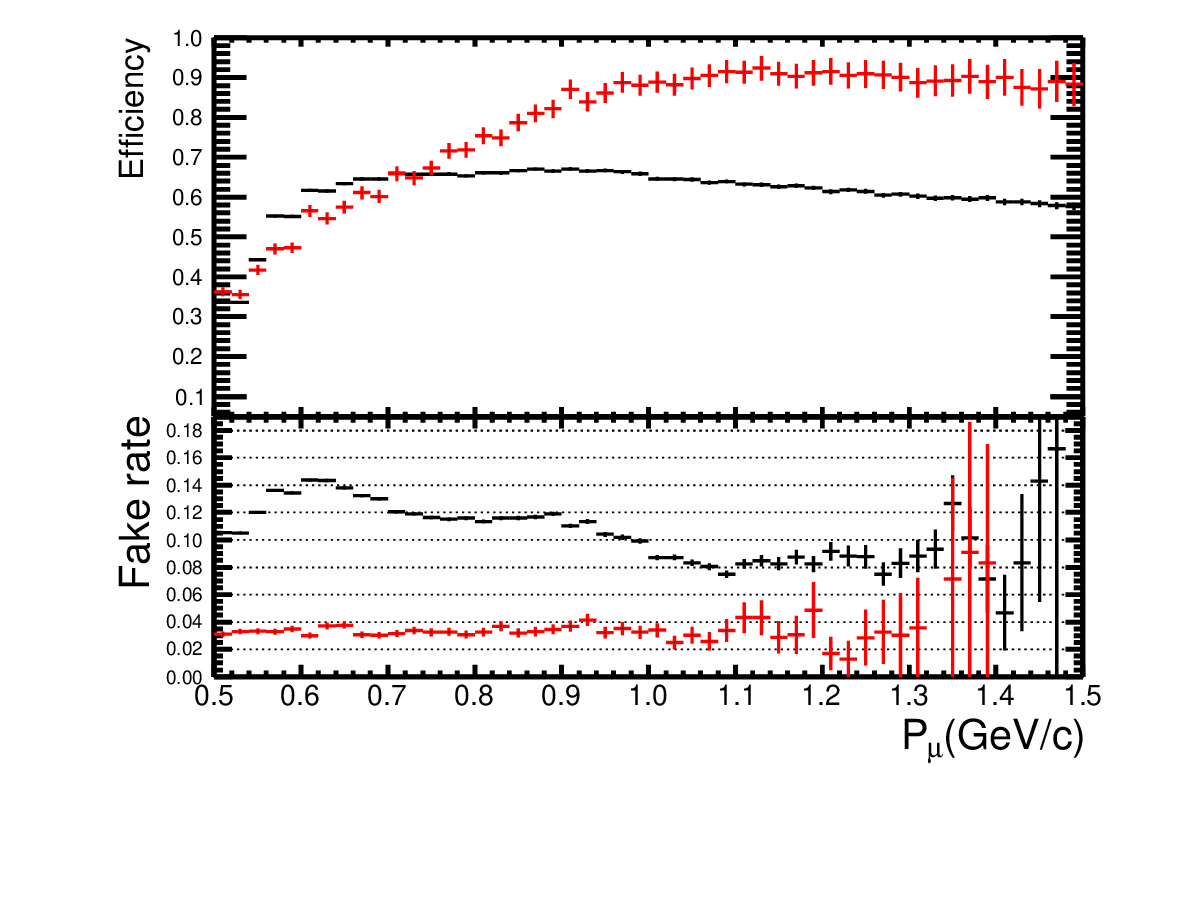}
\end{overpic}
\end{center}
\caption{Detection efficiency of muon and the mis-identification rate of $\pi/\mu$ for two efficiency curves provided by fast simulation}
\label{p_mu_versus_eff}
\end{figure}

\begin{figure*}[htbp]
\begin{center}
\begin{overpic}[width=5.5cm,angle=0]{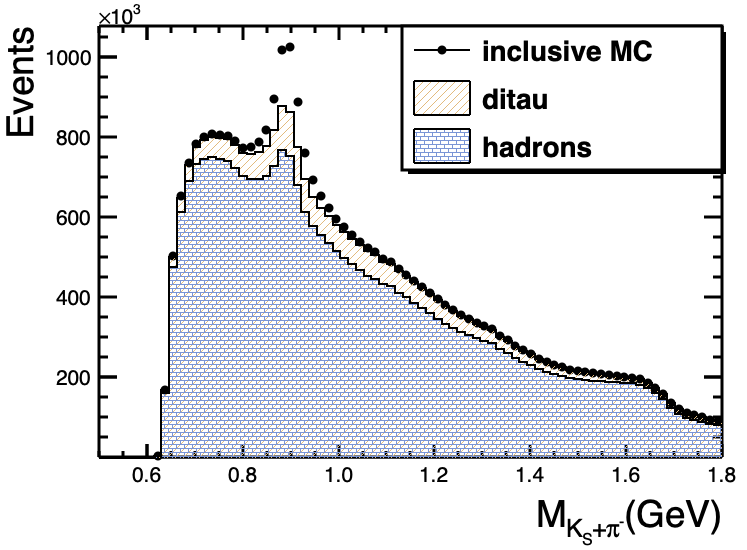}
\put(20,60){\small{(a)}}
\end{overpic}
\begin{overpic}[width=5.5cm,angle=0]{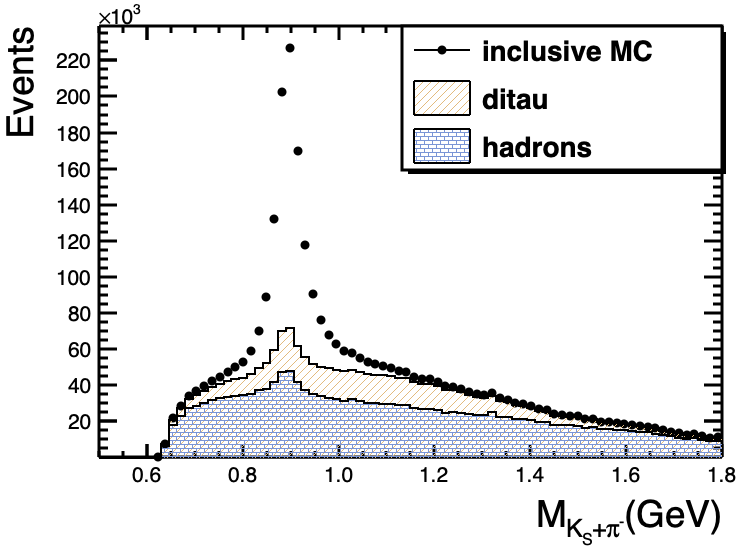}
\put(22,60){\small{(b)}}
\end{overpic}
\begin{overpic}[width=5.5cm,angle=0]{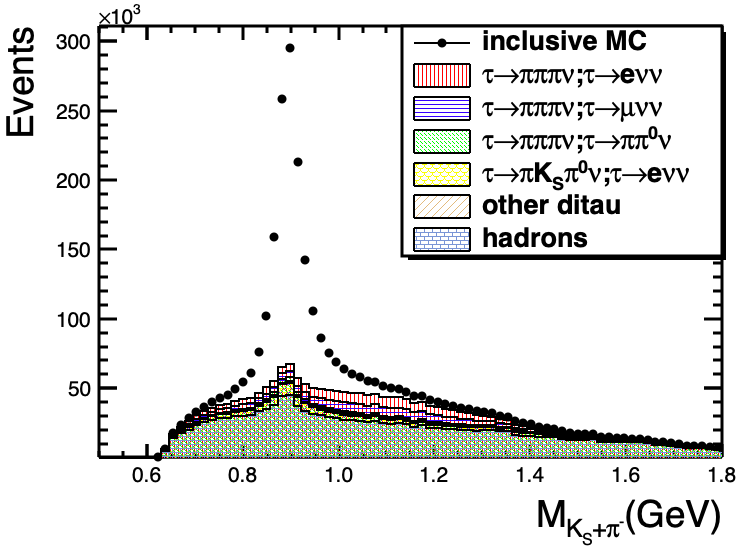}
\put(22,60){\small{(c)}}
\end{overpic}
\end{center}
\caption{Invariant mass spectra of $K_{S}\pi^{-}$ with combined $e$-tag and $\mu$-tag. (a) Distribution without detector optimization or likelihood requirement $y_{L}$ (b) Distribution with likelihood requirement but no detector optimization (c) Distribution with likelihood requirement and detector optimization. MC samples are normalized to 1~ab$^{-1}$ integrated luminosity.}
\label{mass1}
\end{figure*}

After $K_{S}$ selection, a stricter vertex requirement is performed on other charged tracks than the pions in $K_{S}$ decay, 
and number of these tracks satisfying the vertex requirement should equals to 2.
The lepton candidate in tag side and the bachelor pion in signal side will be further selected.
For electron candidates, $E/p$ is required to be larger than 0.8, where $E$ is the deposited energy in the EMC and $p$ is the
momentum of the charged track.
In fast simulation, there are two efficiency curves provided for muon identification, one is from the responses of reference detector~\cite{BESIII},
{\it e.g.} a mixing requirement with $P(\mu)>0.001$, $P(\mu)>P(K)$ and $P(\mu)>P(e)$ in the PID system, where $P(X)$ is the probability of a candidate identified as particle $X$, $0.1<E<0.3$~GeV in EMC, and layer requirement in MUD. 
The efficiency curve is represented as black dots in  Fig.~\ref{p_mu_versus_eff}.
The second one is the response from a preliminary R\&D results of MUD detector at STCF,
where three kinds of $\pi/\mu$ separation levels are provided, to be 3\%, 1.7\% and 1\%, according to different requirements on $\pi/\mu$ responses.
The efficiency curve of muon identification of STCF MUD is also shown in Fig.~\ref{p_mu_versus_eff} in red dots under 3\% $\pi/\mu$ mis-identification requirement. 
After the selection of lepton candidates, the remaining track is identified as pion if it satisfies
 $P(\pi)>P(K)$ and $P(\pi)>P(p)$.
Finally, a signal event is required to have one lepton, one pion and one $K_{S}$.

To suppress the background from $\tau^{-}\to K_{S}\pi^{-}(\geq1 \pi^{0})\nu_{\tau}$ processes,
 $\pi^{0}$ is reconstructed  by selecting
two good photons
and the event is rejected if the invariant mass of any two photons locate in the region between 0.12 and 0.15~GeV/$c^{2}$.
The good photon candidates are selected with the efficiency sampled from fast simulation.

After above selection, the distribution of $K_{S}\pi^{-}$ mass spectrum from generic MC is shown in Fig.~\ref{mass1}(a), where there are huge
backgrounds from
other $\tau$ decay channels in $e^{+}e^{-}\to\tau^{+}\tau^{-}$ excluding signal process, 
and the hadronic final states containing multiple $\pi$ final states. The selection efficiency of
signal and suppressing rate of backgrounds with above selection criteria are listed in Table~\ref{efficiency1}.

 \begin{figure}[htbp]
\begin{center}
\begin{overpic}[width=4.2cm,height=3cm, angle=0]{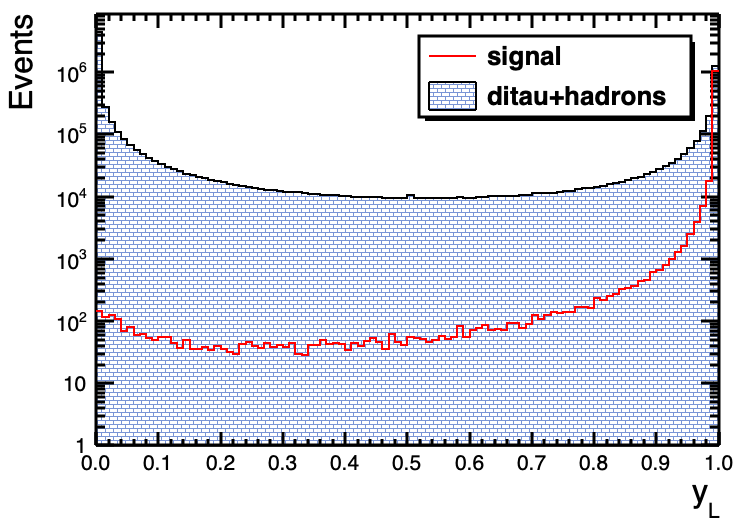}
\put(15,60){\small{(a)}}
\end{overpic}
\begin{overpic}[width=4.2cm,height=3cm, angle=0]{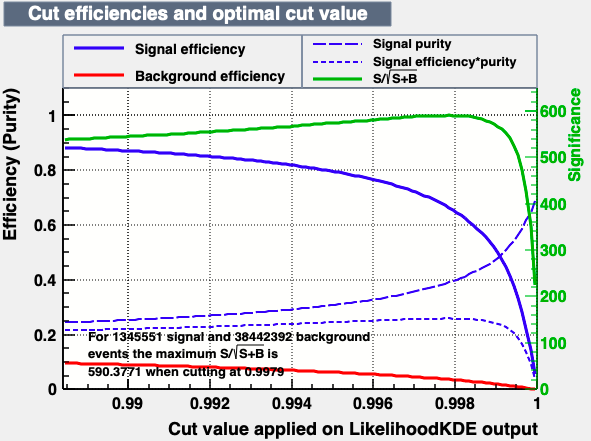}
\put(0,60){\small{(b)}}
\end{overpic}
\end{center}
\caption{(a) Training and test samples distributed. The x-axis denotes $y_{L}$ and y-axis which is set log scale denotes the entries. The signal labeled in figure is the process $\tau^{-}\to K_{S}\pi^{-}\nu_{\tau}$. The backgrounds are $e^{+}e^{-}\to\tau^{+}\tau^{-}$ with $\tau$ decays excluding the signal processes and the hadronic final states. (b) The figure-of-merit for the likelihood selection. MC samples are normalized to 1~ab$^{-1}$ integrated luminosity. }
\label{likelihood1}
\end{figure}

\begin{table*}[htbp]
\centering
\caption{Selection efficiencies for signal and background processes, with neither detector optimization nor likelihood requirement $\varepsilon_{\rm raw}$, with no detector optimization but applying likelihood requirement $\varepsilon_{\rm y_{L}}$, with both detector optimization and likelihood requirement $\varepsilon_{\rm y_{L}\&optimization}$.The errors are statistical only while for background processes, the statistical uncertainties 
are too small and can be neglected.}
\label{efficiency1}
\begin{tabular} {c|ccc }
\hline\hline
~~~\textrm{Process}~~~~~ & ~~~~~~$\varepsilon_{\rm raw}$~\%~~~~~~~& ~~~~~~~~$\varepsilon_{\rm y_{L}}$~\%~~~~ &~~~~~~$\varepsilon_{\rm y_{L}\&optimization}$~\%~~~~~~\\
\hline
$e^{+}e^{-}\to \tau^{+}\tau^{-}$, $\tau^{+}\to l^{+}\nu_{l}\bar{\nu}_{\tau}, \tau^{-}\to K_{S}\pi^{-}\nu_{\tau}$   &$22.80\pm0.02$   & $22.22\pm0.02$&$32.28\pm0.02$\\
$e^{+}e^{-}\to \tau^{+}\tau^{-}$(excluding signal process) & $0.182$  & ~~$0.0403$ & ~~$0.0422$\\
hadronic final states& $0.304$ & ~~$0.0211$ & ~~$0.0179$\\
\hline\hline
\end{tabular}
\end{table*}

To further suppress the background events, a likelihood ratio $y_{L}$ is used defined by
\begin{equation}
y_{L}(\vec{x}) = \frac{L_{S}(\vec{x})}{L_{S}(\vec{x})+L_{B}(\vec{x})},
\end{equation}
where $L_{S}$ and $L_{B}$ are the likelihood function for signal and background event, respectively. $\vec{x}$ is a set of variables used for likelihood. Each likelihood function $L_{S/B}$ is the product of the probability density function(PDF) of the input variables defined by
\begin{equation}
L_{S/B} = \prod^{n_{var}}_{k=1}P_{S/B,k}(x_{k}),
\end{equation}
where $P_{S/B,k}$ is the signal/background PDF of the $k$-$th$ input variable $x_{k}$. For this analysis, the set of variables $\vec{x}$ includes  number of neutral clusters,  momenta of decay products of $K_{S}$, decay length of $K_{S}$,
mass spectrum of  $K_{S}$, $\chi^{2}$ of  $K_{S}$ from secondary vertex fit, $E/p$ ratio of electron, momentum of $\mu$, cosine of the polar angle of $\mu$,  momenta of $\pi$ and $K_{S}\pi$.
The likelihood ratio $y_{L}(\vec{x})$ from these variables between signal and backgrounds are shown in Fig.~\ref{likelihood1} as well as the figure-of-merit.
The classifier requirement on $y_{L}(\vec{x})$ is determined by optimizing the figure-of-merit,
to be 0.9979.
The mass spectrum of $K_{S}\pi^{-}$ after above selections is
shown in Fig.~\ref{mass1}(b), and it is obvious that after likelihood requirements, the background level has been significantly suppressed.
The selection efficiencies for signal and background processes with likelihood requirement are also shown in Table ~\ref{efficiency1}.

\begin{figure}[htbp]
\begin{center}
\begin{overpic}[width=4.2cm,height=3cm, angle=0]{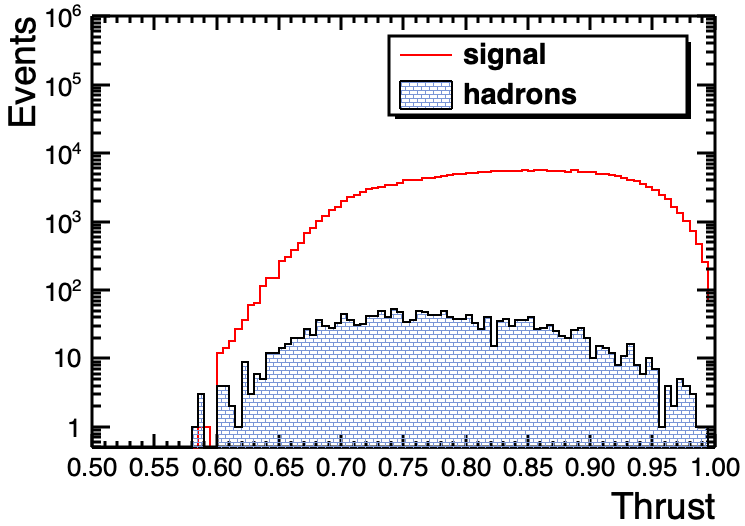}
\put(15,60){\small{(a)}}
\end{overpic}
\begin{overpic}[width=4.2cm,height=3cm, angle=0]{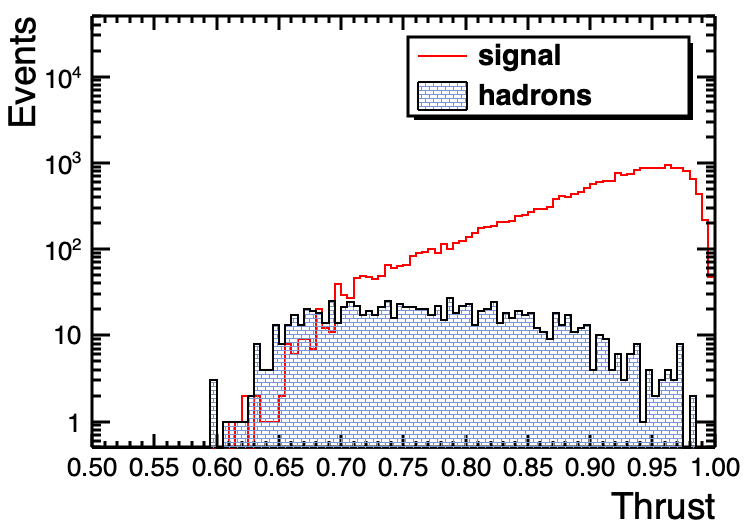}
\put(15,60){\small{(b)}}
\end{overpic}
\end{center}
\caption{The distribution of thrust $T$ (log scale in y-axis) at c.m.energy (a) $\sqrt{s}=4.26$~GeV and (b) $\sqrt{s}=7.0$~GeV, solid line 
in red shows signal process and shaded area in blue shows hadronic processes. 
 MC samples are normalized to 1~ab$^{-1}$ integrated luminosity.}
\label{thrust}
\end{figure}

The background events survived from above selection criteria come from hadronic final states or $e^{+}e^{-}\to\tau^{+}\tau^{-}$
decay products, where $e^{+}e^{-}\to\tau^{+}\tau^{-}$ is mainly from the final processes with one additional $\pi^{0}$,
or multiple pion final states.
At higher c.m.~energies, the thrust $T$ can be used to separate the hadronic final states from signal process, defined as
\begin{equation}
T\equiv max_{\overrightarrow{n}}\frac{\sum_i |\overrightarrow{n}\cdot\overrightarrow{p}_{i}|}{\sum_i |\overrightarrow{p}_{i}|},
\end{equation}
where $\overrightarrow{p}_{i}$ is the 3-momenta
of the final state particles, $\overrightarrow{n}$ is a 3-vector with unit norm. Finding the  $\overrightarrow{n}$ in any direction of space makes the 
thrust $T$ maximum. The typical distribution of thrust $T$ at
c.m.~energy $\sqrt{s}=4.26$ and 7.0~GeV are shown in Fig.~\ref{thrust} between signal and hadronic processes.
It is found that,  the thrust $T$ can provide a good distinguish between
signal and hadronic background at high c.m.~energies  such as $\sqrt{s}=7.0$~GeV.
At lower c.m.~energies, it is hard to distinguish the signal from background due to
insufficient boost. However, with a good
description of the hadronic final states from {\sc LundArlw}, these background can be well described with MC simulation.

\section{\boldmath OPTIMIZATION OF DETECTOR RESPONSE}

After above selection criteria, the signal process is selected with an efficiency of 22.22\%, where the main loss of the efficiency
comes from the effects of track selection, reconstruction of $K_{S}$, and the particle identification of leptons. These
effects correspond to the sub-detectors of inner track, the PID system as well as the MUD.
By studying the signal-to-background ratios in the selection of signal process with variation of sub-detector's responses,
the requirement of detector design can be optimized accordingly.
With the interfaces provided by fast simulation software, four kinds of detector responses can be studied which will be introduced in this section.

~{\it a.Tracking efficiency}
The tracking efficiency in fast simulation is characterized by two dimensions:
transverse momentum $P_{T}$ and polar angle $\cos\theta$, which are correlated with the level of track bending and the hit
positions of tracks in the tracker system.
For low-momentum track
($P_{T}<0.2$~GeV/c), it is difficult to reconstruct efficiently 
due to stronger electromagnetic multiple scattering, electric field leakage, energy loss {\it etc.}.
However, with different technique of inner track design at STCF, or with advanced track finding algorithm, 
the reconstruction efficiency of low-momentum track can be improved.

\begin{figure}[hbtp]
\begin{center}
\begin{overpic}[width=5.5cm,angle=0]{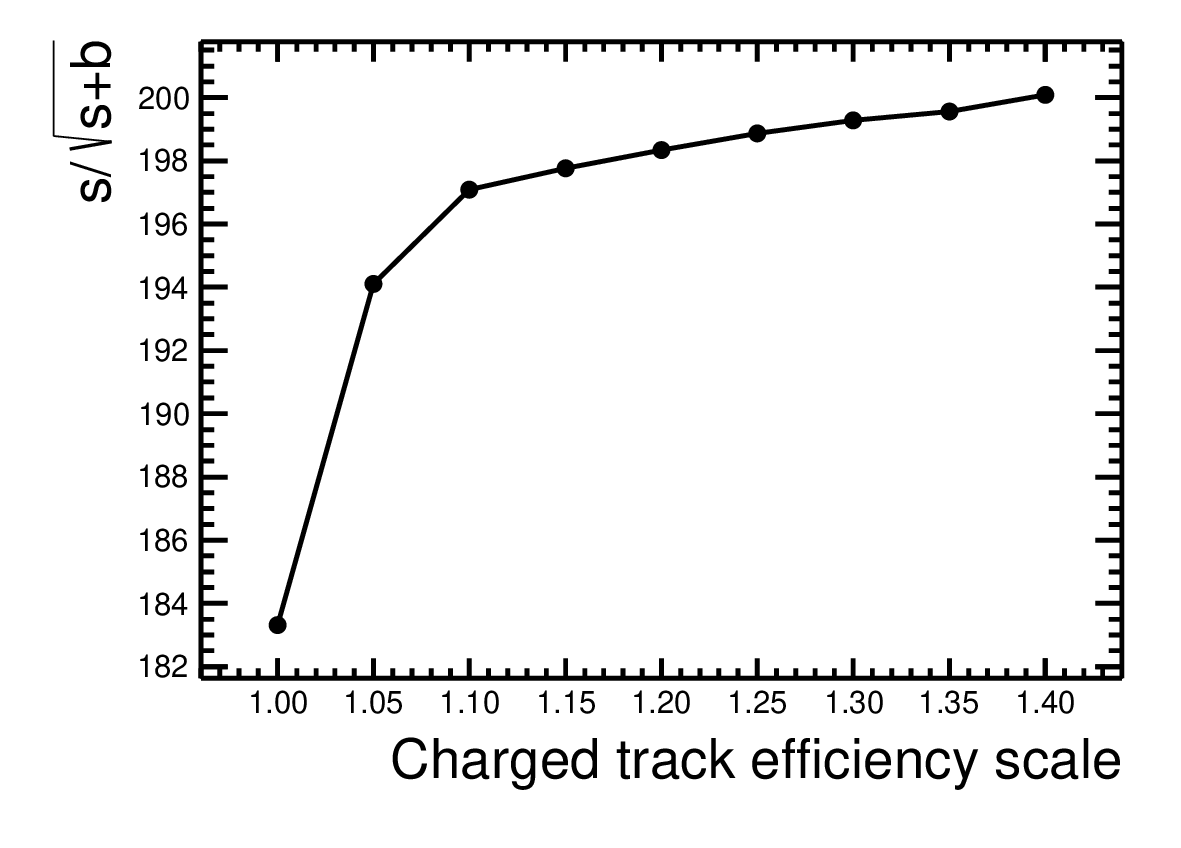}
\end{overpic}
\end{center}
\caption{The figure-of-merit for optimization the tracking efficiency for low-momentum tracks.
}
\label{trackeff1}
\end{figure}

In this analysis, the efficiency is scaled in the fast simulation, with a ratio from 1 to 1.4 for low-momentum tracks,
which indicates the increasing of tracking efficiency by a factor of 10\% to 40\%.
For high-momentum tracks or tracks with efficiency approximate to 100\%, their tracking efficiency keep the same.
The figure-of-merit for each scale factor of low-momentum tracking efficiency is shown in Fig.~\ref{trackeff1},
defined by  $\frac{S}{\sqrt{S+B}}$, where  $S$ denotes signal yields of $\tau^{-}\to K_{S}\pi^{-}\nu_{\tau}$, and $B$ denotes the backgrounds,
both normalized to 1~ab$^{-1}$ integrated luminosity.
From Fig.~\ref{trackeff1}, it is found that the efficiency can be significantly improved with an optimization factor of 1.1 to 1.2.

In fast simulation, the default tracking efficiency is sampled from the performance of the helium-gas-based cylindrical main drift chamber~\cite{fastsimu}.
There are several alternative options for the tracker technique, especially for inner track, such as the silicon pixel detector~\cite{silicon} and the micro-pattern gaseous detector~\cite{gaseous}. From the current R\&D activities, the tracking efficiency for low-momentum track is expected to fulfill the request from our analysis with these techniques at STCF.

~{\it b. Momentum/spatial resolution} The resolution of momentum and spatial can also be optimized in fast simulation with proper scaling factors.
A better resolution will improve the significance of signal with the improvement of resolution in $K_{S}\pi^{-}$ mass spectrum.
The effects of momentum and position resolutions on $K_{S}$ reconstruction  efficiency
is also studied, and it is found that the spatial resolution is more sensitive than that of the momentum resolution for $K_{S}$ reconstruction.
From the reconstruction method for a charged track, these two resolutions are related with the same sources, which are the position resolution of a single wire in $xy$ plane and $z$ direction, and the multiple scattering.
The micro-pattern gaseous detector is allowed to achieve excellent position accuracy~\cite{gaseous}. Besides, a MC study on the
silicon pixel detector shows that the momentum resolution and tracking efficiency can be significantly improved due to its good spatial
resolution~\cite{silicon2}.

~{\it c. $\pi/\mu$ separation}
As discussed before, there are two efficiency curves provided for $\mu$ identification.
The default muon identification in the fast simulation has a large $\pi/\mu$ mis-identification rate
as shown with the black points in Fig.~\ref{p_mu_versus_eff}.
As the fast simulation provides the function for optimizing the $\pi/\mu$ mis-identification,
we can vary the mis-identification rate for $\pi/\mu$  based on the provided efficiency curves.
The signal-to-background ratio can be significantly improved with better $\pi/\mu$ mid-identification rate.
Besides, the $\pi/\mu$ separation power can be significantly improved with a hybrid  design for MUD~\cite{fastsimu},
and the efficiency curves are shown in Fig.~\ref{p_mu_versus_eff} with red points under the requirement of $\pi/\mu$ mis-identification to be 3\%.
Three kinds of mis-identification rates are provided by MUD at STCF, with different efficiency for $\mu$, respectively.
Table~\ref{eff_mupi} summarizes of the selection efficiency for these three efficiency responses,
where one can find, with the $\pi/\mu$ mis-identification at 3\%, the detector efficiency for signal process is the largest,
this is because a moderate selection on the performance at MUD is applied with this requirement. 
Moreover, at STCF, the Ring Imaging CHerenkov detector (RICH) can be used to identify muon from pion at low-momentum.

\begin{table}[htbp!]
\centering
\footnotesize
\caption{ Efficiency of selecting $\mu$ and final seletion efficiency using a  hybrid design for MUD at STCF with different $\pi/\mu$ fake rate (FR) requirements.}
\label{eff_mupi}
\begin{tabular} {c c c c}
\hline\hline
~~~FR/$\%$~~~&~~~~Eff. $\mu/\%$~~~&~~~Eff. bkg$/\%$~~~~&~~~~Final eff.$/\%$~~~~\\
\hline
3&$76.80\pm0.02$& 0.385 &$32.82\pm0.02$\\
1.7&$66.79\pm0.02$&  0.375  &$31.40\pm0.02$\\
1&$58.63\pm0.03$& 0.370   &$30.20\pm0.02$\\
\hline\hline
\end{tabular}
\end{table}

~{\it d. Position/energy resolution for photon}
In this analysis, processes including $\pi^{0}$ are suppressed by vetoing to the events  $\pi^{0}$ reconstructed.
The position and energy resolution for neutral tracks are optimized with the fast simulation.
It is found that the current resolutions provided in fast simulation, which is 
 6~mm for position resolution and 2.5$\%$ for energy resolution of 1~GeV photon, can satisfy the physical requirement for this analysis.

From above study, a set of optimization factors for sub-detector responses concerning this analysis is provided:
an improvement of tracking efficiency by a factor of $10\sim20\%$ for low-momentum tracks;
a good momentum/spatial resolution for charged track;
a good muon identification with a mis-identification rate of $\pi/\mu$ to be less than 3\%;
 and a good position/energy resolution for photons.
With these optimization factors applied,
the selection efficiency for signal process is improved  from 22.22\% to 32.82\%.
Moreover, the invariant mass of $K_{S}\pi^{-}$  are shown in Fig.~\ref{mass1}(c) after optimization.
The background level is significantly suppressed,
where the dominant $\tau$ pair decay background events are $\tau^{-}\to\pi^{+}\pi^{-}\pi^{-}\nu_{\tau}$ and $\tau^{-}\to K_{S}\pi^{-}\pi^{0}\nu_{\tau}$.
The optimized selection efficiency and the background level are summarized in Table~\ref{efficiency1}.
The detailed selection efficiency for each step are listed in Table~\ref{cutflow},
where the improvement of the selection efficiency comparing to the MC without optimization is given.
\begin{table}[ht]
\centering
\caption{
Cut flow for each selection criteria for signal process, where $\varepsilon$ is the overall efficiency, $\varepsilon_{\rm rela}$
is the relative efficiency for each criteria, and $\Delta\varepsilon_{\rm rela}$ is the relatively improvement of optimized
detector response comparing to the original one.
}
\label{cutflow}
\begin{tabular}{c c c c}
\hline\hline
~~~\textbf{Selection}~~~& ~~~$\varepsilon$ (\%)~~~   &  ~~~$\varepsilon_{\rm rela}$ (\%)~~~ & $\Delta\varepsilon_{\rm rela}$ (\%)~~ \\
\hline
$N_{\rm charged\_track}$&$~~69.75\pm0.02$&$69.75$&7.52~~~~\\
select $K_{S}$&$~~54.34\pm0.03$&$77.90$&$13.10~~~~$\\
lepton ID &$~~37.89\pm0.03$&$87.31$&19.13~~~~\\
$\pi$ ID &$~~36.64\pm0.03$&$96.70$&2.66~~~~\\
veto $\pi^{0}$&$~~36.64\pm0.03$&$99.99$&-~~~~\\
$K_{S}$ flight length&$~~33.45\pm0.02$&$91.38$&0.27~~~~\\
likelihood method&$~~32.82\pm0.02$&$98.10$&0.52~~~~\\
\hline
Total&$~~32.82\pm0.02$&$-$&$-~~~~$\\
\hline\hline
\end{tabular}
\end{table}

\section{\boldmath STATISTICAL ANALYSIS}

\begin{figure}[htbp]
\begin{center}
\begin{overpic}[width=4.2cm,height=3cm, angle=0]{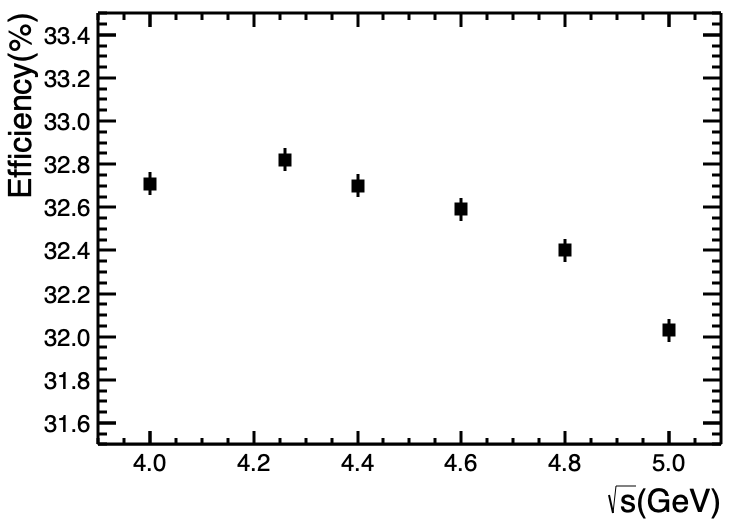}
\put(20,60){\small{(a)}}
\end{overpic}
\begin{overpic}[width=4.2cm,height=3cm, angle=0]{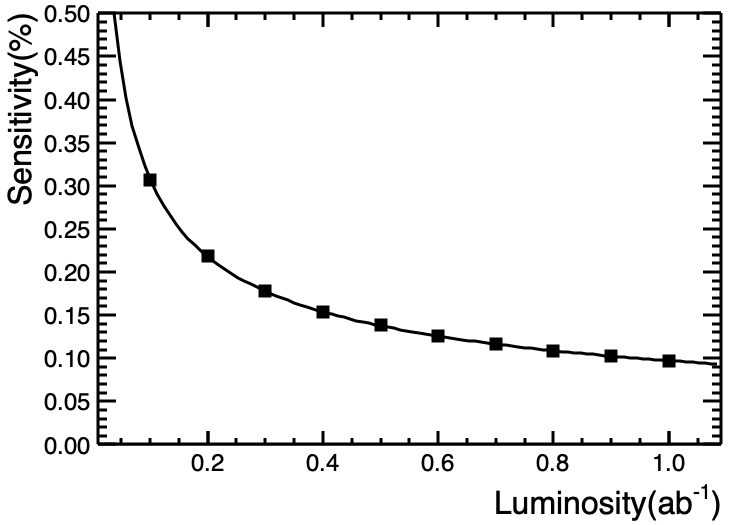}
\put(20,60){\small{(b)}}
\end{overpic}
\end{center}
\caption{ (a) Selection efficiencies for signal process at different c.m.~energies from $\sqrt{s}=4.0$~GeV to 5.0~GeV.
(b) The CPV sensitivity of $\tau^{-}\to K_{S}\pi^{-}\nu_{\tau}$ under different integrated luminosity $L$, which follows $1/\sqrt{L}$ extrapolation.}
\label{energy_eff}
\end{figure}

With above selection criteria and optimization procedure, 
the number of signal events from $\tau^{-}$ and $\tau^{+}$ decay with generic MC normalized to 1~ab$^{-1}$ integrated luminosity are obtained by
fit the $K_{S}\pi$ invariant mass with RooFit tool,
where signal can be parameterized by the function as shown in Eq.~(\ref{equation3}) and background is described with simulated background.
The efficiency corrected numbers for $\tau^{-}\to K_{S}\pi^{-}\nu_{\tau}$ and $\tau^{+}\to K_{S}\pi^{+}\bar{\nu}_{\tau}$ are
$3681017\pm5034$ and $3681127\pm5091$, respectively.  It shows a good consistency with input values.
The statistical sensitively of CPV with decay rate can be calculated according to Eq.~(\ref{ACPTh}), to be $9.7\times10^{-4}$.
Since the cross section of $e^{+}e^{-}\to\tau^{+}\tau^{-}$ is around 3.5~nb in the energy region from $\sqrt{s}=4.0$ to 5.0~GeV,
we studied the selection efficiency for the signal process in this energy region, as shown in Fig.~\ref{energy_eff}. 
The efficiencies in this energy region vary from 32.82\% to 32.03\% with a relative difference of less than 3\%.  
Therefore, the statistics of signal process is creasing linearly with more data collected. 
As it is not background free in the event selection, the sensitively of CPV is proportional to the $1/\sqrt{L}$
as shown in Fig.~\ref{energy_eff}(b) where different amount of MC samples from 0.1~ab$^{-1}$ to 1.0~ab$^{-1}$ are applied for the study.
Finally, we can conclude that, with 10~ab$^{-1}$ integrated luminosity collected at STCF from $\sqrt{s}=4.0$ to 5.0~GeV, 
the sensitively of the CPV for the process $\tau^{-}\to K_{S}\pi^{-}\nu_{\tau}$ is  $3.1\times10^{-4}$.

Though the analysis is performed with purely MC simulation, we need to discuss possible systematic uncertainty 
which may bias the decay-rate asymmetry measurement between $\tau^{-}\to K_{S}\pi^{-}\nu_{\tau}$ and $\tau^{+}\to K_{S}\pi^{+}\bar{\nu}_{\tau}$.
The uncertainty sources will be introduced and estimated in following:
i) the detection asymmetry for charged particles, which can be studied from control sample of $\tau^{-}\to \pi^{+}\pi^{-}\pi^{-}\nu_{\tau}$.
The difference from detector can be corrected by comparing the asymmetry between MC simulation and data, and the remain uncertainty is 
related with the statistics uncertainty of the control sample. Since the branching fraction of the control sample is over one magnitude
larger than signal process, the uncertainty will be significantly smaller than the statistics uncertainty of $\tau^{-}\to K_{S}\pi^{-}\nu_{\tau}$.
ii) uncertainty related with selection criteria, which can be studied by varying the selection criteria. By performing the Barlow test~\cite{barlow}, 
any bias from selection will be studied and corrected until the systematic accuracy matches the statistical precision.
iii) uncertainty from MC generator. In this analysis, no CPV is modeled and the output asymmetry in MC is about $3\times10^{-5}$ which can be neglected.
iv) uncertainty from background contamination. In this analysis, there is no decay-rate asymmetry observed in background, and the uncertainty can be tested by applying  different criteria on the background between data and MC simulation.
v) the different nuclear-interaction cross sections of the $K^{0}$ and $\bar{K}^{0}$ meson. In Ref.~\cite{CPbabar}, a correction to the
asymmetry accounting from $K^{0}$ and $\bar{K}^{0}$ mesons with material in the detector is applied  by taking the $K^{\pm}$ 
nucleon cross sections as a analogy under the assumption of isospin invariance~\cite{k0asymmetry}. We take an uncertainty of 0.01\% 
for this uncertainty similar to that from Ref.~\cite{CPbabar} ignoring the materia difference at STCF.
As a consequence of above discussion, the systematic uncertainty for decay-rate asymmetry study of $\tau^{-}\to K_{S}\pi^{-}\nu_{\tau}$
will be at the same level of its statistical uncertainty.

As introduced before, the sensitivity of
CPV in the processes of  $\tau^{-}\to K_{S}\pi^{-}\nu_{\tau}$ at BelleII experiment is expected to be $1/\sqrt{70}$ of the current precision of Belle experiment
with a 50~ab$^{-1}$ data sample collected~\cite{CPbelleii}, which is better than the sensitivity obtained at STCF with one-year data taking, but 
comparable with  that at STCF with ten-years data taking. The study of  $\tau^{-}\to K_{S}\pi^{-}\nu_{\tau}$ at B-factory benefits from
a larger luminosity, and a good separation of two hemispheres from $\tau^{+}$ and $\tau^{-}$ decay. However, at STCF, the cross section of
$e^{+}e^{-}\to\tau^{+}\tau^{-}$ is the largest and the selection efficiency for the signal process is better with high detection/PID efficiency
for low-momentum tracks. When the c.m.~energy goes up, {\it e.g.} at $\sqrt{s}=7$~GeV, the two hemispheres from $\tau^{+}$ and $\tau^{-}$ decays
can also be well separated, as shown in Fig.~\ref{thrust}.  Therefore, it is of great potential to study the CPV and relevant physics program
of the process  $\tau^{-}\to K_{S}\pi^{-}\nu_{\tau}$ at STCF.

\section{\boldmath SUMMARY AND PROSPECT}
In this paper, the sensitivity of decay-rate asymmetry in $\tau^{-}\to K_{S}\pi^{-}\nu_{\tau}$ decays is studied at $\sqrt{s}=4.26$~GeV
with 1~ab$^{-1}$ inclusive MC.
Benefit from the largest cross section of 3.5~nb for $e^{+}e^{-}\to\tau^{+}\tau^{-}$ at $\sqrt{s}=4.26$~GeV
 and the optimized efficiency with the fast simulation,
the statistical sensitivity of CPV is 
$9.7\times10^{-4}$, which is 2.3 times improved comparing to that of BaBar~\cite{CPbabar}.
With 10~ab$^{-1}$ luminosity collected at STCF, the sensitivity is expected to be at a level of $3.1\times10^{-4}$, which is comparable to
the uncertainty of theoretical prediction in the SM. A new round measurement of $A_{CP}$ defined by the decay rate difference is important and
desirable, c.f.~\eqref{expcp}, which will certainly shed light on the existence of the NP signal. Several theoretical models
discussed in Sec.~2 will be reexamined. The CPV arose by the exchange of charged Higgs boson has also been explored in experiments,
which is defined by the difference between angular observables, as a helpful complement. 
With such precision of $10^{-4}$, the remeasurement
will be pursued. In the meanwhile, other mentioned prospects (form factors, $|V_{us}|$ extraction etc.) could be furbished. \\

\section*{\boldmath ACKNOWLEDGEMENTS}
The authors thank the supercomputing center of USTC and Hefei Comprehensive National Science Center for their strong support. This work is supported by the Double First-Class university project foundation of USTC,
and the National Natural Science Foundation of China under Projects No.11625523.
The author XWK is supported by the National Natural Science Foundation of China under Project No.11805012, and the Fundamental Research Funds for the
Central Universities.\\



\begin{thebibliography}{99}
\bibitem{CKM1}  N.~Cabibbo, Unitary Symmetry and Leptonic Decays, Phys.\ Rev.\ Lett.\  {\bf 10}, 531 (1963) doi: 10.1103/PhysRevLett.10.531.
\bibitem{CKM2} M.~Kobayashi and T.~Maskawa, CP Violation in the Renormalizable Theory of Weak Interaction, Prog.\ Theor.\ Phys.\  {\bf 49}, 652 (1973) doi: 10.1143/PTP.49.652.
\bibitem{kaoncp1}  H.~Burkhardt {\it et al.} [NA31 Collaboration], First Evidence for Direct CP Violation, Phys.\ Lett.\ B {\bf 206}, 169 (1988) doi: 10.1016/0370-2693(88)91282-8.
\bibitem{kaoncp2} V.~Fanti {\it et al.} [NA48 Collaboration], A New measurement of direct CP violation in two pion decays of the neutral kaon, Phys.\ Lett.\ B {\bf 465}, 335 (1999) doi: 10.1016/S0370-2693(99)01030-8.
\bibitem{kaoncp3} A.~Alavi-Harati {\it et al.} [KTeV Collaboration], Observation of direct CP violation in $K_{S,L}\to\pi\pi$ decays, Phys.\ Rev.\ Lett.\  {\bf 83}, 22 (1999) doi: 10.1103/PhysRevLett.83.22.


\bibitem{Bcp1} B.~Aubert {\it et al.} [BaBar Collaboration], Observation of direct CP violation in $B^{0}\to K^{+}\pi^{-}$ decays, Phys.\ Rev.\ Lett.\  {\bf 93}, 131801 (2004) doi: 10.1103/PhysRevLett.93.131801.
\bibitem{Bcp2}  Y.~Chao {\it et al.} [Belle Collaboration], Evidence for direct CP violation in $B^{0}\to K^{+}\pi^{-}$ decays, Phys.\ Rev.\ Lett.\  {\bf 93}, 191802 (2004) doi: 10.1103/PhysRevLett.93.191802.
\bibitem{Bcp3} A.~Poluektov {\it et al.} [Belle Collaboration], Evidence for direct CP violation in the decay $B\to D^{*}K, D\to K_{S}\pi^{+}\pi^{-}$ and measurement of the CKM phase phi3, Phys.\ Rev.\ D {\bf 81}, 112002 (2010) doi: 10.1103/PhysRevD.81.112002.
\bibitem{Bcp4} P.~del Amo Sanchez {\it et al.} [BaBar Collaboration], Measurement of CP observables in $B^{+-}\to D_{CP} K^{+-}$ decays and constraints on the CKM angle $\gamma$, Phys.\ Rev.\ D {\bf 82}, 072004 (2010) doi: 10.1103/PhysRevD.82.072004.
\bibitem{Bcp5} R.~Aaij {\it et al.} [LHCb Collaboration], Observation of CP violation in $B^{+}\to DK^{+}$ decays ,Phys.\ Lett.\ B {\bf 712}, 203 (2012) Erratum: [Phys.\ Lett.\ B {\bf 713}, 351 (2012)] doi:	10.1016/j.physletb.2012.05.060.
\bibitem{Bcp6} R.~Aaij {\it et al.} [LHCb Collaboration], First observation of CP violation in the decays of $B^{0}_{s}$ mesons, Phys.\ Rev.\ Lett.\  {\bf 110}, 221601 (2013) doi: 10.1103/PhysRevLett.110.221601.

\bibitem{charmcp} R.~Aaij {\it et al.} [LHCb Collaboration], Observation of CP Violation in Charm Decays ,Phys.\ Rev.\ Lett.\  {\bf 122}, 211803 (2019) doi: 10.1103/PhysRevLett.122.211803.

\bibitem{matterasy1}  A.~D.~Sakharov, Violation of CP Invariance, C asymmetry, and baryon asymmetry of the universe, Pisma Zh.\ Eksp.\ Teor.\ Fiz.\  {\bf 5}, 32 (1967) doi: 10.1070/PU1991v034n05ABEH002497.
\bibitem{matterasy2}  M.~E.~Shaposhnikov, Possible Appearance of the Baryon Asymmetry of the Universe in an Electroweak Theory, JETP Lett.\  {\bf 44}  465 (1986).

\bibitem{extendSM1}  H.~P.~Nilles, Supersymmetry, Supergravity and Particle Physics, Phys.\ Rept.\  {\bf 110}, 110 (1984) doi: 10.1016/0370-1573(84)90008-5.
\bibitem{extendSM2}  H.~E.~Haber and G.~L.~Kane, The Search for Supersymmetry: Probing Physics Beyond the Standard Model, Phys.\ Rept.\  {\bf 117}, 75 (1985) doi: 10.1016/0370-1573(85)90051-1.

\bibitem{Yanagida}
W.~Buchmuller, R.~D.~Peccei and T.~Yanagida, Leptogenesis as the origin of matter,
Ann. Rev. Nucl. Part. Sci. \textbf{55} 311-355 (2005) doi: 10.1146/annurev.nucl.55.090704.151558.

\bibitem{taucp1} Y.~S.~Tsai, Search for new mechanism of CP violation through tau decay and semileptonic decay of hadrons, Nucl.\ Phys.\ Proc.\ Suppl.\  {\bf 55C}, 293 (1997) doi: 10.1016/S0920-5632(97)00226-0.
\bibitem{taucp2} I.~I.~Bigi, Probing CP Violation in $\tau^{-}\to\nu(K\pi/K2\pi/3K/K3\pi)^{-}$ Decays, arXiv:1204.5817v2 [hep-ph].
\bibitem{taucp3} K.~Kiers, CP violation in hadronic $\tau$ decays, Nucl.\ Phys.\ Proc.\ Suppl.\  {\bf 253-255}, 95-98 (2014) doi: 10.1016/j.nuclphysbps.2014.09.023.

\bibitem{BellePhysicsBook}
E.~Kou \textit{et al.} [Belle-II], The Belle II Physics Book,
PTEP \textbf{2019}, 123C01 (2019) doi: 10.1093/ptep/ptz106, 10.1093/ptep/ptaa008 (erratum).

\bibitem{Pich}
A.~Pich, Precision Tau Physics,
Prog. Part. Nucl. Phys. \textbf{75}, 41-85 (2014) doi: 10.1016/j.ppnp.2013.11.002.

\bibitem{belleii}
E.~Kou \textit{et al.} [Belle-II],
PTEP \textbf{2019}, no.12, 123C01 (2019)
[erratum: PTEP \textbf{2020}, no.2, 029201 (2020)]
doi:10.1093/ptep/ptz106
[arXiv:1808.10567 [hep-ex]].

\bibitem{stcf}
H.-p. Peng, High Intensity Electron Positron Accelerator (HIEPA), Super Tau Charm Facility (STCF) in China, talk at Charm2018, Novosibirsk, Russia, May 21 - 25, 2018.
 Q.~Luo and D.~Xu, Progress on Preliminary Conceptual study of HIEPA, a super tau-charm factory in China, talk at the 9th International Particle Accelerator Conference (IPAC 2018), held in Vancouver, British Columbia, Canada, April 29 - May 4, 2018.

\bibitem{tsai}
Y.~S.~Tsai,
Production of polarized tau pairs and tests of CP violation using polarized e+- colliders near threshold,
Phys. Rev. D \textbf{51}, 3172-3181 (1995)
doi:10.1103/PhysRevD.51.3172
[arXiv:hep-ph/9410265 [hep-ph]].


\bibitem{taucp5}I.~I.~Bigi and A.~I.~Sanda, A 'Known' CP asymmetry in tau decays, Phys.\ Lett.\ B {\bf 625}, 47 (2005) doi: 10.1016/j.physletb.2005.08.033.
\bibitem{taucp6}Y.~Grossman and Y.~Nir, CP Violation in $\tau^{\pm}\to\pi^{\pm}K_{S}\nu$ and $D^{\pm}\to\pi^{\pm}K_{S}$: The Importance of $K_{S}-K_{L}$ Interference, JHEP {\bf 1204}, 002 (2012) doi: 10.1007/JHEP04(2012)002.

\bibitem{CPbabar} J.~P.~Lees {\it et al.} [BaBar Collaboration], Search for CP Violation in the Decay $\tau^{-}\to\pi^{-}K^{0}_{S}(\ge0\pi^{0})\nu_{\tau}$, Phys.\ Rev.\ D {\bf 85}, 031102 (2012) doi: 10.1103/PhysRevD.85.031102, 10.1103/PhysRevD.85.099904.

\bibitem{tensor1}
H.~Z.~Devi, L.~Dhargyal and N.~Sinha, Can the observed CP asymmetry in $\tau \to K\pi\nu_{\tau}$ be due to nonstandard tensor interactions?,
Phys. Rev. D \textbf{90}, 013016 (2014) doi: 10.1103/PhysRevD.90.013016.

\bibitem{tensor2}
L.~Dhargyal, Full angular spectrum analysis of tensor current contribution to $A_{cp}(\tau \rightarrow K_{s} \pi \nu_{\tau})$,
LHEP \textbf{1}, 9-14 (2018) doi: 10.31526/LHEP.3.2018.03.

\bibitem{Martin}
V.~Cirigliano, A.~Crivellin and M.~Hoferichter, No-go theorem for nonstandard explanations of the $\tau\to K_S\pi\nu_\tau$ CP asymmetry,
Phys. Rev. Lett. \textbf{120}, 141803 (2018) doi: 10.1103/PhysRevLett.120.141803.


\bibitem{RoigPRD}
J.~Rendon, P.~Roig and G.~Toledo, Effective-field theory analysis of the $\tau^{-}\rightarrow (K \pi)^{-}\nu_{\tau}$ decays,
Phys. Rev. D \textbf{99}, 093005 (2019) doi: 10.1103/PhysRevD.99.093005.

\bibitem{XinQiangLi}
F.~Z.~Chen, X.~Q.~Li, Y.~D.~Yang and X.~Zhang, CP asymmetry in $\tau\to K_S\pi\nu_\tau$ decays within the Standard Model and beyond,
Phys. Rev. D \textbf{100}, 113006 (2019) doi: 10.1103/PhysRevD.100.113006.

\bibitem{Roy}
A.~Dighe, S.~Ghosh, G.~Kumar and T.~S.~Roy, Tensors for tending to tensions in $ \tau $ decays,
[arXiv:1902.09561 [hep-ph]].


\bibitem{2HDM}
N.~Mileo, K.~Kiers and A.~Szynkman, Probing sensitivity to charged scalars through partial differential widths: $\tau\rightarrow K\pi\pi\nu_{\tau}$ decays,
Phys. Rev. D \textbf{91}, 073006 (2015) doi: 10.1103/PhysRevD.91.073006.

\bibitem{Kimura}
D.~Kimura, K.~Y.~Lee and T.~Morozumi, The Form factors of $\tau \to K \pi(\eta) \nu$ and the predictions for CP violation beyond the standard model,
PTEP \textbf{2013} 053B03 (2013) doi: 10.1093/ptep/ptt013, 10.1093/ptep/ptu107 (erratum), 10.1093/ptep/ptt084 (erratum).

\bibitem{KangTodd}
X.~W.~Kang and H.~B.~Li, Study of CP violation in D$\to$VV decay at BESIII,''
Phys. Lett. B \textbf{684} 137-140 (2010),
Int. J. Mod. Phys. A \textbf{26} 2523-2535 (2011) doi: 10.1016/j.physletb.2010.01.024,

\bibitem{Kang1}
I.~I.~Bigi, X.~W.~Kang and H.~B.~Li,
CP Asymmetries in Strange Baryon Decays,
Chin. Phys. C \textbf{42}, 013101 (2018) doi: 10.1088/1674-1137/42/1/013101.

\bibitem{Kang2}
X.~D.~Shi, X.~W.~Kang, I.~Bigi, W.~P.~Wang and H.~P.~Peng, Prospects for CP and P violation in $\Lambda_{c}^+$ decays at Super Tau Charm Facility,
Phys. Rev. D \textbf{100}, 113002 (2019) doi: 10.1103/PhysRevD.100.113002.

\bibitem{CPcleo} G.~Bonvicini {\it et al.} [CLEO Collaboration], Search for CP violation in tau$\to$K pi tau-neutrino decays, Phys.\ Rev.\ Lett.\  {\bf 88}, 111803 (2002) doi: 10.1103/PhysRevLett.88.111803.

\bibitem{CPbelle}  M.~Bischofberger {\it et al.} [Belle Collaboration], Search for CP violation in $\tau \to K^0_S \pi \nu_\tau$ decays at Belle, Phys.\ Rev.\ Lett.\  {\bf 107}, 131801 (2011) doi: 10.1103/PhysRevLett.107.131801.

\bibitem{theocp} J.~H.~Kuhn and E.~Mirkes, Structure functions in tau decays, Z.\ Phys.\ C {\bf 56}, 661 (1992) doi: 10.1007/BF01474741 (publication), 10.1007/BF01571299 (erratum).

\bibitem{CPbelleii}
F.~Z.~Chen, X.~Q.~Li and Y.~D.~Yang,
$CP$ asymmetry in the angular distribution of $\tau\to K_S\pi\nu_\tau$ decays,
JHEP \textbf{05}, 151 (2020)
doi:10.1007/JHEP05(2020)151
[arXiv:2003.05735 [hep-ph]].

\bibitem{Bernard}
V.~Bernard, First determination of $f_+(0) |V_{us}|$ from a combined analysis of $\tau\to K\pi \nu_\tau$ decay and $\pi K$ scattering with constraints from $K_{\ell3}$ decays,
JHEP \textbf{06}, 082 (2014) doi: 10.1007/JHEP06(2014)082.


\bibitem{Roig2014}
R.~Escribano, S.~Gonzalez-Solis, M.~Jamin and P.~Roig, Combined analysis of the decays $\tau^{-} \to K_{S} \pi^{-} \nu_{\tau}$ and $\tau^{-} \to K^{-} \eta\nu_{\tau}$,
JHEP \textbf{09}, 042 (2014) doi: 10.1007/JHEP09(2014)042.

\bibitem{Oller}
M.~Jamin, J.~A.~Oller and A.~Pich, Scalar K pi form factor and light quark masses,
Phys. Rev. D \textbf{74}, 074009 (2006) doi: 10.1103/PhysRevD.74.074009.

\bibitem{Belle2007}
D.~Epifanov \textit{et al.} [Belle Collaboration], Study of tau$\to$ K(S) pi- nu(tau) decay at Belle,
Phys. Lett. B \textbf{654}, 65-73 (2007) doi: 10.1016/j.physletb.2007.08.045.

\bibitem{Vusforkspi}
V.~Bernard, D.~R.~Boito and E.~Passemar, Dispersive representation of the scalar and vector $\rm{K}\pi$ form factors for $\tau \to K \pi \nu_\tau$ and $K_{\ell3}$ decays, Nucl.Phys.B Proc.Suppl. \textbf{218} 140-145 (2011) doi: 10.1016/j.nuclphysbps.2011.06.024.
\bibitem{Vusforkspi2}
M.~Antonelli, V.~Cirigliano and A.~Lusiani, Predicting the $\tau$ strange branching ratios and implications for $V_{us}$, JHEP \textbf{10} 070 (2013) doi: 10.1007/JHEP10(2013)070.


\color{black}
\bibitem{fastsimu} X.~D.~Shi, X.~R.~Zhou, X.~S.~Qin and H.~P.~Peng, A fast simulation package for STCF detector, arXiv:2011.01654 [physics.ins-det].
\bibitem{babayaga} G.~Balossini {\it et al.}, Matching perturbative and parton shower corrections to Bhabha process at flavour factories, Nucl.\ Phys.\ B {\bf 758}, 227 (2006) doi: 10.1016/j.nuclphysb.2006.09.022.
G.~Balossini {\it et al.}, Photon pair production at flavour factories with per mille accuracy, Phys.\ Lett.\ B {\bf 663}. 209 (2008) doi: 10.1016/j.physletb.2008.04.007.
\bibitem{lund} B.~Andersson and H.~Hu, Few body states in Lund string fragmentation model, hep-ph/9910285.
\bibitem{KKMC} S. Jadach, B. F. Ward and Z. Was, The Precision Monte Carlo event generator K K for two fermion final states in e+ e- collisions, Comput. Phys. Commun. {\bf 130}, 260 (2000) doi: 10.1016/S0010-4655(00)00048-5.
\bibitem{pdg} M.~Tanabashi {\it et al.} [Particle Data Group], Review of Particle Physics, Phys. Rev. D {\bf 98}, 030001 (2018) doi: 10.1103/PhysRevD.98.030001.

\bibitem{belleamp}  D.~Epifanov {\it et al.} [Belle Collaboration], Study of tau$\to$ K(S) pi- nu(tau) decay at Belle, Phys.\ Lett.\ B {\bf 654}, 65 (2007) doi: 10.1016/j.physletb.2007.08.045.

\bibitem{BESIII}
J.~Z.~Bai \textit{et al.} [BES],
The BES detector,
Nucl. Instrum. Meth. A \textbf{344}, 319-334 (1994)
doi:10.1016/0168-9002(94)90081-7.


%


%



\bibitem{silicon}
R.~Turchetta, J.~D.~Berst, B.~Casadei, G.~Claus, C.~Colledani, W.~Dulinski, Y.~Hu, D.~Husson, J.~P.~Le Normand and J.~L.~Riester, \textit{et al.}
A monolithic active pixel sensor for charged particle tracking and imaging using standard VLSI CMOS technology,
Nucl. Instrum. Meth. A \textbf{458}, 677-689 (2001)
doi:10.1016/S0168-9002(00)00893-7.

\bibitem{gaseous}
L.~I.~Shekhtman,
Micro-pattern gaseous detectors,
Nucl. Instrum. Meth. A \textbf{494}, 128-141 (2002)
doi:10.1016/S0168-9002(02)01456-0.

\bibitem{silicon2}
Q.~Xiu, M.~Dong, W.~Li, H.~Liu, Q.~Ma, Q.~Ouyang, Z.~Qin, L.~Wang, L.~Wu and Y.~Yuan, \textit{et al.}
Study of the Tracking Method and Expected Performance of the Silicon Pixel Inner Tracker Applied in BESIII,
[arXiv:1510.08558 [physics.ins-det]].


\bibitem{barlow}
R.~Barlow,
Systematic errors: Facts and fictions,
[arXiv:hep-ex/0207026 [hep-ex]].


\bibitem{k0asymmetry}
B.~R.~Ko, E.~Won, B.~Golob and P.~Pakhlov,
Effect of nuclear interactions of neutral kaons on CP asymmetry measurements,
Phys. Rev. D \textbf{84}, 111501 (2011)
doi:10.1103/PhysRevD.84.111501
[arXiv:1006.1938 [hep-ex]].
\end{thebibliography}
\end{document}